# Tunable Photonic Radiofrequency Filter with An Ultra-high Out-Of-Band Rejection

Peixuan Li, Xihua Zou, *Member, IEEE*, Wei Pan, Lianshan Yan, *Senior Member, IEEE* and Shilong Pan, *Senior Member, IEEE*

*Abstract*—As radiofrequency filtering plays a vital role in electromagnetic devices and systems, recently photonic techniques have been intensively studied to implement radiofrequency filters to harness wide frequency coverage, large instantaneous bandwidth, low frequency-dependent loss, flexible tunability and strong immunity to electromagnetic interference. However, one crucial challenge facing the photonic radiofrequency filter (PRF) is the less impressive out-of-band rejection. Here, to the best of our knowledge, we demonstrate a tunable PRF with a record out-of-band rejection of 80 dB, which is 3 dB higher than the maximum value (~77 dB) reported so far, when incorporating highly selective polarization control and large narrow-band amplification enabled by stimulated Brillouin scattering effect. In particular, this record rejection is arduous to be achieved for a narrow passband (e.g., a few megahertz) and a high finesse in a PRF. Moreover, the proposed PRF is an active one capable of providing negligible insertion loss and even signal gain. Tunable central frequency ranging from 2.1 to 6.1 GHz is also demonstrated. The proposed PRF will provide an ultra-high noise or clutter suppression for harsh electromagnetic scenarios, particularly when room-temperature implementation and remote distribution are needed.

## I. INTRODUCTION

High-performance tunable radiofrequency (RF) filters are highly required in electromagnetic devices and systems, such as future 5G wireless communication and beyond, cognitive radio, agile radar, new-generation electronic warfare and deep-space astronomy [1]-[3]. A large number of architectures and materials for designing and fabricating tunable electronic RF filters have been reported [4], [5], to achieve desired specifications for diverse applications and scenarios. In particular, the out-of-band rejection, a key figure of merit of an RF filter, has gained considerable attention for fully suppressing noises, clutters and jamming signals in harsh electromagnetic scenarios. As an example, an out-of-band rejection as high as 80 dB or even 90 dB can be achieved [6]-[11] by using superconductor materials or elements working at an ultra-low temperature of 70 K or dual-mode cavities in the electrical domain. Nevertheless, wide frequency coverage and flexible tuning over a large fractional bandwidth are still challenging for these traditional electronic RF filters, due to the limited speed and bandwidth arising from electronic bottleneck.

Microwave photonics, which brings together the RF engineering and optoelectronics [12], [13], takes the intrinsic advantages of photonic technologies to provide basic units or devices [14-18] and to enrich or enhance the functions of microwave systems [19]-[35] that are complex or even not directly possible in the electrical domain. Typically, significant functions enabled by microwave photonics include the generation [19]-[21], processing [22], [23], measurement or detection [24]-[28], and distribution [29]-[36] of microwave signals.

Over the past years, the photonic RF filter (PRF) defined as RF filter assisted by photonics, has shown superior performance in terms of large instantaneous bandwidth, wide frequency coverage, fast frequency tunability and simple reconfigurability [37]-[55], with respective to electronic counterparts. However, regarding the out-of-band rejection or signal selectivity, as a key figure of merit for RF filters, the PRFs are generally less competitive, showing an out-of-band rejection no more than 40 dB in most reports. Here the out-of-band rejection is defined as the difference between the peak (the maximum) of the noise floor and the transmission peak. There are several exceptional examples with high out-of-band rejection (see more details in Table I). PRFs based on multi-tap delay line architectures were demonstrated with a high out-of-band rejection of ~70 dB and rapid tenability [37], [38], while providing periodic spectral response. Assisted by cascaded optical resonators, the PRF was able to offer a high out-of-band rejection over 70 dB [41], [42]. When using a Fabry-Perot microresonator, an outstanding rejection of ~77 dB [42] can be obtained for a 650-MHz passband. To reduce the passband further to tens or a few megahertz for achieving a high processing resolution or a high finesse, the out-of-band rejection will be degraded by 10 dB or more [42], [44].

In general, however, the maximum out-of-band rejection reported so far for PRFs is still not comparable with the value of state-of-the-art electronic filters based on superconductor materials or elements operating at ultra-low temperature (e.g., 70 K in [6]). Therefore it is extremely challenging to realize PRF with competitively high out-of-band rejection, particularly for the need of fine filtering resolution or high finesse.

Manuscript received November, 2016. This work was supported in part by the National "863" Project of China under grant 2015AA016903 and by the National Natural Science Foundation of China under grant 61378008. X. Zou was also supported by the Fellowship from the Alexander von Humboldt Foundation, Germany.

P. Li, X. Zou. W. Pan, and L. Yan are with the Center for Information Photonics and Communications, School of Information Science and Technology, Southwest Jiaotong University, Chengdu 610031, China (zouxihua@swjtu.edu.cn; wpan@swjtu.edu.cn).

S. Pan is with Key Laboratory of Radar Imaging and Microwave Photonics, Ministry of Education, Nanjing University of Aeronautics and Astronautics, Nanjing 210016, China.

Currently, stimulated Brillouin scattering (SBS) activated in optical fiber [56] or on-chip device [57] is used to perform optical and microwave signal processing, such as sensing [58], slow light [59] and stored light [60], true time reversal [61] and polarization control [62], with high resolution owing to the ultra-narrow gain or/and loss bandwidth of SBS. In particular, SBS can be exploited as a powerful tool to perform RF filtering [43]-[51] with high out-of-band rejection. A notch PRF with an out-of-band rejection over 60 dB was achieved based on the SBS in a segment of fiber [47]. By using a Brillouin-active photonic crystal waveguide, an out-of-band rejection of 70 dB was obtained for the PRF with a passband as narrow as 3.15 MHz [44]. However, this filter lacks tunability in the central frequency, compared with other tunable PRFs with relatively lower rejection [46]. Although these PRFs based on SBS above are still not competitive with the superconductor-based RF filters having ultra-high out-of-band rejection, they suggest great potential for reducing the gap.

Here, we propose a PRF through the combination of the highly selective polarization control and large narrowband amplification enabled by SBS, demonstrating a record high out-of-band rejection up to 80 dB. In the proposed PRF, the incoming RF signal is applied to an electro-optic polarization modulator (PolM) to externally modulate an optical carrier. Under the double-sideband (DSB) modulation, two optical sidebands that are out of phase are generated with the same state of polarization (SOP). But the SOP of the two optical sidebands and that of the optical carrier are orthogonal [63]. Thanks to the unique feature of polarization modulation, the out-of-band noise induced by the phase and amplitude imbalances of two sidebands can be relieved due to the orthogonal SOPs between the two sidebands and optical carrier. Furthermore, undesirable signals (e.g., noises or clutters) resulting from the imperfect orthogonality of SOPs, can be canceled as the two optical sidebands are out of phase by 180°. The SBS amplification is also used to selectively rotate the SOP of one of the two sidebands and to boost the amplitude of the same sideband. Therefore, a narrow passband with an ultra-high out-of-band rejection can be expected. In the experiments, the spectral response of the proposed PRF is measured to achieve a record high out-of-band rejection over 80 dB, which is 3 dB higher than the maximum value ever reported for PRFs [42]. In particular, this record rejection is achieved for a narrow bandwidth of 7.7 MHz and a high finesse of 844, which might be 10 dB or more higher than that for a passband of a few megahertz [37], [42], [44]. Tunable central frequency is also available for the proposed PRF.

## II. PRINCIPLE

The principle of the PRF is illustrated in Fig. 1. An optical carrier is modulated by the incoming RF signal at the PolM. As shown in Fig. 1(a), two optical sidebands that are 180° out of phase are generated with the same SOP which is orthogonal to that of the optical carrier, under the condition of small-signal modulation. First of all, when the SBS processing is switched off, only a direct-current signal can be detected from the polarization modulation after opto-electronic conversion and thus the spectral response is null in microwave bands. Next, the SBS effect is activated to pull the SOP and to boost the amplitude of one of the two optical sidebands generated by applying the incoming RF signal to the PolM. The SOP of the optical sideband can be pulled to identically align with that of the optical carrier, as shown in Fig. 1(b). Due to the changes in the SOP and in the amplitude profile of the polarization-modulated optical signal induced by the SBS effect, a conversion from polarization modulation to intensity modulation is realized and consequently the incoming RF signal can be recovered. Furthermore, the SOP control and amplitude amplification are performed with high selectivity within an ultra-narrow frequency range that is equal to the nature SBS bandwidth. As a result, only the RF signal with one of corresponding optical sidebands generated through polarization modulation falling inside the SBS bandwidth can be recovered, while other undesired RF components or clutters are removed. Therefore, a single-passband PRF can be realized to achieve an extremely high out-of-band rejection. Here the central frequencies ($\omega_{gain}$ and $\Omega_{filter}$) of the SBS gain and of the passband of the PRF can be derived as

$$\omega_{gain} = \omega_c + \Omega_p - \Omega_B, \qquad (1)$$

$$\Omega_{filter} = \Omega_p - \Omega_B, \qquad (2)$$

where $\omega_c$ is the angular frequency of the optical carrier, $\Omega_p$ is the angular frequency difference between the pump signal and optical carrier, and $\Omega_B$ is the Brillouin frequency shift. From (2), it is clear that the central frequency of the bandpass PRF can be precisely tuned by changing $\Omega_p$.

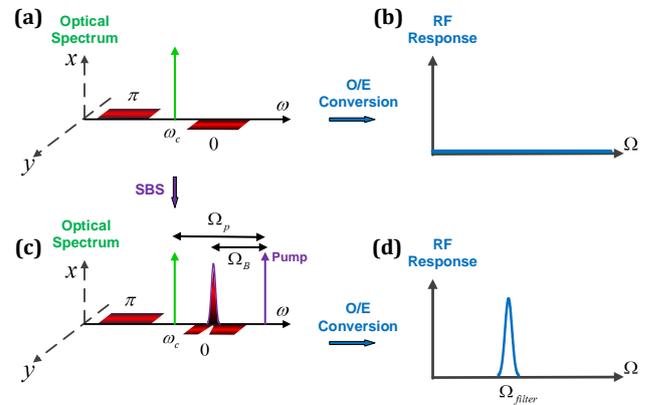

Fig. 1. Illustration of the operation principle of the PRF. Optical spectra of the polarization modulation when switching off (a) or on (c) the SBS processing. RF spectral responses of the PRF when switching off (b) or on (d) SBS processing. $x$ and $y$ represent two orthogonal polarization axes.

### A. Polarization Modulation

The RF signal here is applied to a commercial AlGaAs/GaAs PolM which is capable of supporting two orthogonal eigen modes. Applying an electrical field to the PolM will induce two equal but out-of-phase changes in the refractive indices along two orthogonal polarization directions of the two eigen modes [64]. When the optical carrier linearly polarized at 45° with respective to one principal polarization axis of the PolM, is modulated by the incoming RF signal, the output field of the PolM can be written as [63]

$$\vec{E}_{PolM} = \begin{bmatrix} E_x \\ E_y \end{bmatrix} \propto \frac{E_{in}}{\sqrt{2}} \begin{bmatrix} \cos(\omega_c t + \gamma \cos\Omega t) \\ \cos(\omega_c t - \gamma \cos\Omega t) \end{bmatrix}, \quad (3)$$

where $x$ and $y$ represent two orthogonal polarization axes which are aligned with the axes of the PolM, $E_x$ and $E_y$ are the decomposed components of output fields along $x$ and $y$ polarization axes, $E_{in}$ is the amplitude of the optical carrier, and $\gamma$ is the modulation depth. For simplicity, we assume a sinusoidal RF signal with an angular frequency of $\Omega$. For a linear and small-signal modulation, as schematically illustrated in Fig. 2(a) left, (3) can be simplified as

$$\begin{bmatrix} E_x \\ E_y \end{bmatrix} \propto m_1 \begin{bmatrix} \frac{m_0}{m_1}\cos(\omega_c t) + \cos[(\omega_c+\Omega)t + \frac{\pi}{2}] - \cos[(\omega_c-\Omega)t - \frac{\pi}{2}] \\ \frac{m_0}{m_1}\cos(\omega_c t) - \cos[(\omega_c+\Omega)t + \frac{\pi}{2}] + \cos[(\omega_c-\Omega)t - \frac{\pi}{2}] \end{bmatrix}, \quad (4)$$

where $m_0$ and $m_1$ denote the amplitudes of the optical carrier and optical sidebands. Equation (4) demonstrates that the decomposed components for the two optical sidebands at $\omega_1 = \omega_c - \Omega$ and $\omega_2 = \omega_c + \Omega$ along $x$ and $y$ axes are out of phase by 180°, while the decomposed components of the optical carrier along $x$ and $y$ axes are in phase. Thus, as shown in Fig. 2(a) right, the synthesized field vectors for the optical carrier and two optical sidebands can be derived as

$$\vec{E}_{PolM} \propto \left\{ \begin{bmatrix} 1 \\ 1 \end{bmatrix} \cos(\omega_c t) + \begin{bmatrix} 1 \\ -1 \end{bmatrix} [\cos(\omega_c t + \Omega t + \frac{\pi}{2}) - \cos(\omega_c t - \Omega t - \frac{\pi}{2})] \right\}. \quad (5)$$

Note that the two generated optical sidebands are with the same SOP, but they are out of phase in analog with the feature of a phase modulation [46]. Moreover, an orthogonal relationship is observed between the SOP of the two optical sidebands and that of the optical carrier, as shown on the right of Fig. 2(a). To confirm these relationships, the SOPs of the optical sidebands and optical carrier are measured by a polarization analyzer (PA) and shown in Fig. 2(b). On the Poincaré sphere, the SOPs of the optical carrier (marked as A) and two optical sidebands (marked as B) are almost orthogonal, despite a slight deviation due to the limited extinction ratio of the PolM. This slight deviation imposes negligible impact on suppressing undesired RF noises or clutters due to the nearly orthogonal SOPs between the optical sidebands and the optical carriers. On the other hand, the two out-of-phase optical sidebands can further relieve this unfavorable impact.

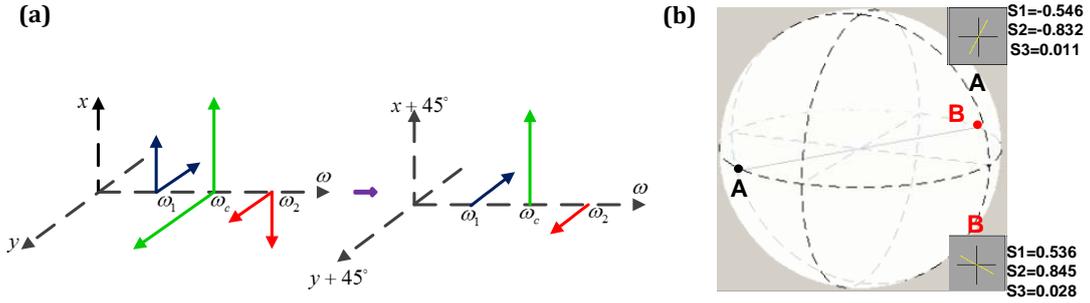

Fig. 2. SOP analysis and measurement for polarization modulation. (a) Schematic illustration of the phase relationship of the two optical sidebands and optical carrier in the polarization modulation (left) and the synthesized field vectors for the optical components of the optical carrier and optical sidebands along two orthogonal principal polarization axes of the PolM. (b) Measured SOPs of the optical carrier (marked as A) and two optical sidebands (marked as B) on the Poincaré sphere. The insets show the measured polarization ellipses and Stokes vector parameters. Here $x$ and $y$ represent two orthogonal polarization axes.

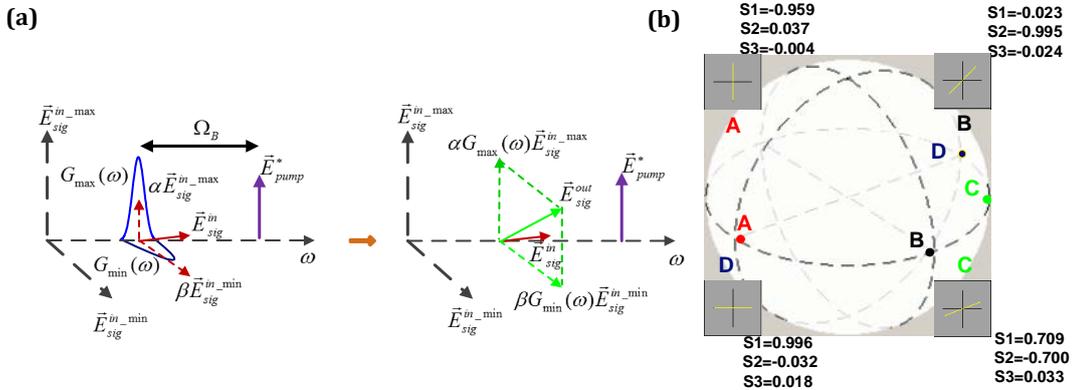

Fig. 3. Polarization control using SBS. (a) Schematic illustration of the polarization pulling process using SBS. (b) Measured SOPs of the unprocessed probe signal (marked as A), the polarization-pulled probe signal after the first-stage SBS (marked as B), the pump signal for the first-stage SBS (marked as C), and the polarization-pulled probe signal after the second-stage SBS (marked as D) on the Poincaré sphere. The insets show the measured polarization ellipses and Stokes vector parameters. The SOP of the probe signal ($\vec{E}_{sig}^{in}$) which falls into the SBS gain bandwidth is pulled towards that of $\vec{E}_{sig}^{in\_max}$

## B. Polarization Control Using SBS Amplification in Fiber

The SBS amplification in standard, randomly birefringent fibers (typically, standard single-mode fiber) is highly polarization-dependent [62], [66]. As seen in Fig. 3(a), when a probe signal ($\vec{E}_{sig}^{in}$) and a pump signal are injected into an optical fiber in opposite directions, an ultra-narrow gain band downshifted by the Brillouin frequency shift, $\Omega_B$, can be generated in the propagation direction of the probe signal. When the probe signal falls inside the SBS gain bandwidth, it can be decomposed into two components along the two orthogonal bases ($\vec{E}_{sig}^{in\_max}$ and $\vec{E}_{sig}^{in\_min}$) that denote the maximum and minimum SBS gains [62]. Then the optical signal at the output end of the fiber is expressed as

$$\vec{E}_{sig}^{out} = \alpha G_{max}(\omega) \vec{E}_{sig}^{in\_max} + \beta G_{min}(\omega) \vec{E}_{sig}^{in\_min}, \quad (6)$$

where $\alpha$ and $\beta$ are the amplitudes of the two orthogonal components of the input probe signal, $G_{max}(\omega)$ and $G_{min}(\omega)$ denote the maximum and minimum values of the SBS-induced gain. Generally, for an undepleted pump, we have $G_{max}(\omega) >> G_{min}(\omega)$ [65] and thus the SOP of the input probe signal can be pulled towards that of $\vec{E}_{sig}^{in\_max}$. Moreover, the SOP of $\vec{E}_{sig}^{in\_max}$ is identical with that of the complex conjugate of the input pump signal ($\vec{E}_{pump}^*$ in Fig. 3(a)). Therefore, the SOP of the probe signal can be flexibly controlled by adjusting the SOPs of the pump signal and the SBS gain.

## III. Experiments for Radiofrequency Filtering

The layout and experimental setup for the proposed PRF are shown in Figs. 4 (a) and (b), respectively. A single-wavelength optical carrier with an ultra-narrow linewidth (<1 kHz) is generated from a laser diode and split into a pump path and a probe path by using a 50:50 optical coupler (OC), one servicing as the probe signal and the other as the pump signal. In the pump path, the pump signal is generated through carrier-suppressed single-sideband (CS-SSB) modulation. Here, a 40-GHz Mach-Zehnder intensity modulator biased at the minimum transmission point to suppress the optical carrier and an optical bandpass filter used to remove the lower sideband are incorporated to perform CS-SSB modulation, when a sinusoidal microwave signal at $\Omega_p$ is applied. Therefore, the frequency of the pump signal is upshifted by $\Omega_p$. The frequency difference, $\Omega_p$, between the pump signal and optical carrier in the probe path can be aligned in a high precision of sub-Hertz by controlling a microwave synthesizer. A two-stage SBS architecture is utilized to provide both highly selective polarization control and large amplitude amplification. The pump signal boosted by an Erbium-doped fiber amplifier (EDFA), is equally split and launched into two 25-km standard single-mode fiber spools via optical circulators. Polarization controllers (PCs) are used to control the SOPs of the pump signals in the two stages to achieve an optimal polarization pulling effect. The low-intensity probe signal is coupled into the two fiber spools in an opposite propagation direction with respective to the pump signal for activating SBS interaction.

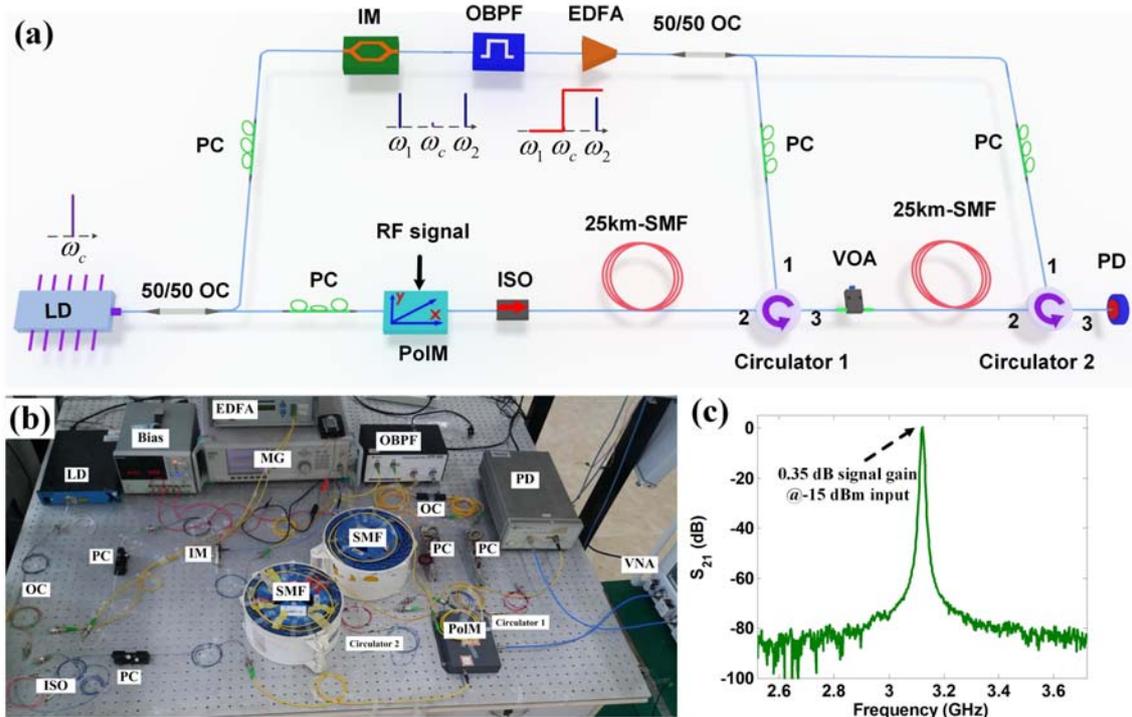

Fig. 4. (a) Layout and (b) photograph of the experimental setup of the proposed PRF. (c) Measured spectral responses ($S_{21}$) with a signal gain of 0.35 dB. LD: laser diode; OC: optical coupler; PC: polarization controller; MG: microwave signal generator; PolM, polarization modulator; ISO: isolator; SMF: single mode fiber; IM: Mach-Zehnder intensity modulator; OBPF: optical bandpass filter; EDFA: Erbium-doped fiber amplifier; VOA: variable optical attenuator; PD: photodetector.

The polarization pulling effect of the two-stage SBS in fibers is firstly verified by analyzing the SOPs, when no RF signal is applied to modulate the optical carrier in the probe path. The pump signal is generated by setting $\Omega_p = \Omega_B = 10.879$ GHz, which is then amplified to 13 dBm and equally split into the two fiber spools to interact with the probe signal. The SOPs of the pump signal and probe signal are measured by the PA and depicted in Fig. 3(b). Without SBS processing on the probe signal, the SOP of the probe signal at port 3 of the first optical circulator (circulator 1) is marked as point A on the Poincaré sphere. Then, the pump signal is switched on to provide SBS interaction and the SOP of the probe signal after the first-stage SBS is marked as point B. Point C indicates the SOP of the complex conjugate of the pump signal, which was measured by switching off the probe path and only the spontaneous SBS is excited to act as a polarization mirror of the pump signal [66]. As shown in Fig. 3(b), the SOP of the probe signal (marked as point A) is pulled towards that of the pump signal and a detuning of 180° in the SOP can be observed on the Poincaré sphere after the two-stage SBS interaction.

After the characterization of the SOPs, an incoming RF signal is applied to the PolM with a bandwidth of 40 GHz and a low half-wave voltage of 3.5 V, to modulate the optical carrier in the probe path to demonstrate the filtering function. The pump signal is amplified to have a power level of 13 dBm and equally split into the two fiber spools. When $\Omega_p$ is specified as 14 GHz, a 2-port vector network analyzer (VNA) is used to measure the spectral response ($S_{21}$) of the proposed PRF. As can be seen from Fig. 4(c), a single passband centered at 3.12 GHz is observed, when the pump signal is switched on. Owing to the SBS gain, the proposed PRF is an active one capable of providing signal gain, rather than insertion loss in conventional PRFs. When the power level of the input signal is set as -15 dBm, a signal gain of 0.35 dB is obtained for the passband.

To show more details, the normalized response (dark blue line) is presented in Fig 5(a). An ultra-high out-of-band rejection up to 80 dB is achieved in the frequency span of 7 GHz, for the passsband with a full-width at half-maximum of 7.7 MHz. A zoom-in view of the passband is also shown in Fig. 5(b). For the purpose of comparison, the spectral response (or the noise floor) is also recorded as the green curve, when the pump signal is switched off. Due to highly selective amplification of the SBS processing, only a slight increase on the noise floor is observed for the proposed PRF, which is considered as a significant contribution to the achievement of a record out-of-band rejection. To highlight the out-of-band rejection achieved in the proposed PRF, a comprehensive collection on current PRFs with an out-of-band rejection greater than 40 dB is shown in Table I. Here the achieved 80-dB rejection is 3 dB greater than the maximum value of 77 dB ever reported for the PRF with a 650-MHz passband [42]. It should be highlighted that the 80-dB rejection is available for a narrow bandwidth of 7.7 MHz and hence a high finesse of 844. In general, it is arduous to achieve an ultra-high rejection for a narrower passband in PRFs, since in the optical domain a high out-of-band rejection is usually accompanied with a wider passband (e.g., DWDM). To reduce the passband to provide a high processing resolution, the out-of-band rejection will be degraded. Consequently, for a passband of tens or a few megahertz, this 80-dB rejection is one order of magnitude or more higher than that reported previously [37], [42], [44], enabling an ultra-high out-of-band rejection under a high processing resolution.

Also, this 80-dB rejection is even comparable with the specification of state-of-the-art electronic RF filters operating at stringent ambient temperature (e.g., 70 K for superconducting materials [6]) and one order of magnitude higher than that of the PRFs with a few megahertz bandwidth.

On the other hand, as predicted in (2), the central frequency of the single passband can be precisely tuned by adjusting $\Omega_p$. Tunable spectral responses centered from 2.1 to 6.1 GHz are obtained with a tuning step of 1 GHz, as shown in Figs. 5 and 6. According to all the measured spectral responses, an ultra-high out-of-band rejection close to 80 dB can be observed. It should be mentioned that, due to the limited bandwidth with high sensitivity and low noise level of the test instruments and transimpedance amplifier use in the experiments, the out-of-band rejection will degrade as the central frequency increases. Thus an ultra-high rejection is unavailable for high frequencies beyond 10 GHz at this moment. In fact, the frequency coverage will be greatly extended if high-performance test instruments are used.

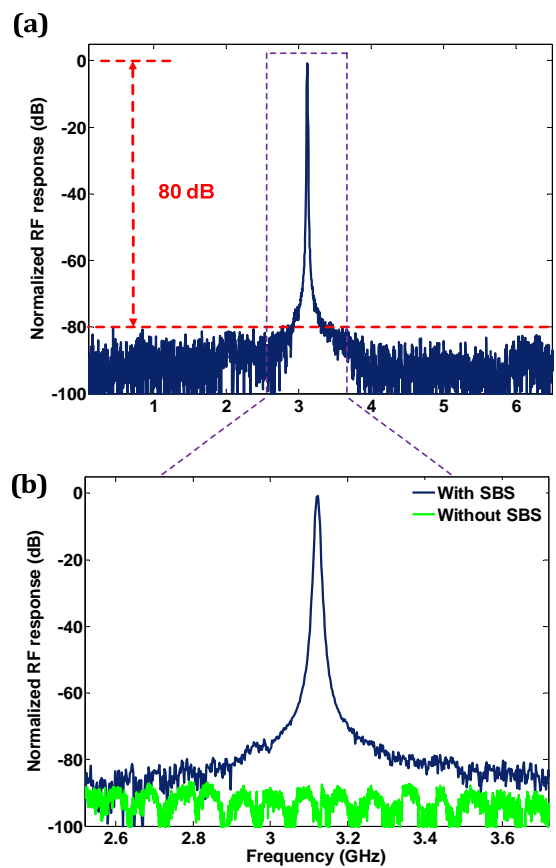

Fig. 5. Measured spectral responses of the proposed PRF. (a) Measured spectral responses centered at 3.1 GHz with (dark blue) and without (green) SBS processing, demonstrating an out-of-band rejection of 80 dB. (b) Zoom-in view of the spectral response within a span of 1.2 GHz.

## IV. DISCUSSION

The proposed PRF is demonstrated by using discrete optoelectronic devices and fiber spools, which might be considered to be bulky and low power efficiency. But this version of the proposed PRF here can be simply incorporated into existing RF signal processing systems by using commercially available and cost-effective components operating at room temperature, with respective to the ultra-low temperature required by superconductive filters [6]. The use of long fiber link also facilitates remote RF interception in electronic warfare and radar. Furthermore, with the rapid development of the photonic integrated circuit (PIC) technology [23, 53], it is expected to develop an integrated version of this PRF with a much smaller footprint and higher power efficiency.

TABLE I
SELECTED PRFs WITH REJECTION LARGER THAN 40 dB

| PRFs (ref.) | Category | Tunability | Fixed frequency or frequency coverage (GHz) | 3-dB Bandwidth (MHz) | Rejection (dB) |
|---|---|---|---|---|---|
| 46 | Notch | Yes | 2-20 | 31 | 42 |
| 40 | Notch | Yes | 2.5-17.5 | 6000-9500 | ~45 |
| 54 | Notch | Yes | 10-40 | — | 48 |
| 45 | Notch | Yes | 14-20 | 98 | 48 |
| 53 | Notch | Yes | 0-50 | — | ~50 |
| 43 | Notch | Yes | 1-30 | 33-88 | ~55 |
| 39 | Notch | Yes | 2-8 | 247-840 | ~60 |
| 47 | Notch | Yes | 1-30 | 10-65 | ~60 |
| 55 | Notch | Yes | 12.4-30.6 | ~12500 | >60 |
| 37 | Periodic passbands | — | 0.14 | ~0.44 | 52/70 |
| 38 | Periodic passbands | Yes | 9.38 | 170-800 | ~61/70 |
| 48,50 | Single passband | Yes | 1.6-2.15 | 250-1000 | 44/46 |
| 51 | Single passband | Yes | ~8.5 | 500-3000 | 45 |
| 52 | Single passband | Yes | 9.68 | — | ~55 |
| 44 | Single passband | NO | 2.93 | 3.15 | 70 |
| 41 | Single passband | — | ~11.5 | 650 | ~70 |
| 42 | Single-passband | — | — | 650 | ~77 |
| **This Article** | **Single passband** | **Yes** | **2.1-6.1** | **7.7** | **80** |

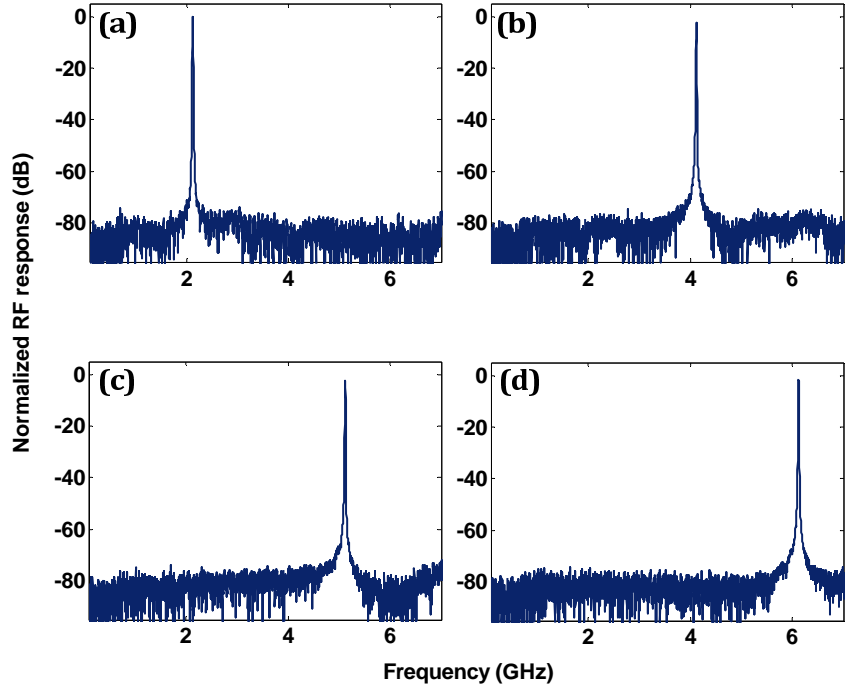

Fig. 6. Demonstration of tunable central frequency. The spectral responses are measured with tunable central frequencies at (a) 2.1 GHz, (b) 4.1 GHz, (c) 5.1 GHz, (d) 6.1 GHz. Combined with the spectral response centered at 3.1 GHz shown in Fig. 5, tunable central frequencies ranging from 2.1 to 6.1 GHz have been verified experimentally.

The PRF is also featured by the capability of processing RF or microwave signal with low power level, making it particularly available for applications in cognitive radio, electronic warfare and deep-space astronomy. There are two reasons behind this capability. First, to perform an effective polarization pulling, an undepleted pump regime is generally assumed [48, 62, 65, 67]. Correspondingly, the power level of the probe signal, which is indirectly associated with that of the incoming RF signal, should be sufficiently low. Secondly, the incoming RF signal with low power level can improve the SBS gain of the proposed system. Taking a 10-dBm pump signal for example, the SBS amplification in a fiber link over 2 km can easily reach the saturated regime when injecting a low-power probe signal (<-40 dBm) [68]. In the saturated regime, the output power of the probe signal is remained as constant and insensitive with the power increase of the probe signal. Therefore, in the saturated regime, the relative SBS gain can be improved by decreasing the power level of the probe signal, while keeping a fixed absolute SBS gain. Accordingly, to achieve high SBS gain and thereby high selectivity, a small-signal condition (<-30 dBm) [48, 51, 72] is generally assumed for the operation of the SBS-based PRF. However, note that an extremely low power level to the probe signal is unlikely to be applicable, since it will induce strong amplified spontaneous emission noise [70, 71].

V. CONCLUSION

We demonstrated a tunable PRF by incorporating highly selective polarization control and large narrow-band amplification enabled by SBS effect. The PRF has a record high out-of-band rejection of 80 dB which is 3 dB higher than the maximum value (77 dB [42]) ever reported for PRFs. In particular, the 80-dB rejection is available for a narrow bandwidth of 7.7 MHz, which is one order of magnitude greater than that reported previously for a bandwidth of a few megahertz in PRFs. The proposed PRF is also capable of providing tunable central frequency from 2.1 to 6.1 GHz while remaining an ultra-high rejection. Such an ultra-high out-of-band rejection allows the proposed PRF to meet the stringent demands on high selectivity of a target RF signal and on high suppression of noises or clutters in diverse applications such as cognitive radio, electronic warfare and deep-space astronomy.

ACKNOWLEDGMENT

The authors would like to thank Prof. Jianping Yao (Microwave Photonics Research Laboratory, University of Ottawa, Canada) for his comments and suggestions to improve the quality of this article.

**Peixuan Li** received the BS degree from Southwest Jiatotong University, China, in 2012. He is currently working toward the Ph.D. degree in the School of Information Science and Technology, Southwest Jiaotong University. His current research interests include microwave photonic processing and generation, fiber communication systems.

**Xihua Zou** is currently a Full Professor in the Center for Information Photonics & Communications, Southwest Jiaotong University, China. Since October 2014, he has been working as a Humboldt Research Fellow in the Institute of Optoelectronics, University of Duisburg-Essen, Germany. He once was a visiting researcher and a joint training Ph.D. student in the Microwave Photonics Research Laboratory, University of Ottawa, Canada, in July and August of 2011 and in 2007-2008, respectively. He was also a visiting scholar in the Ultrafast Optical Processing Research Group, INRS-EMT, Canada, in July and August of 2012. His current interests include microwave photonics, radio over fiber, and optical communications. He has authored or coauthored over 80 academic papers in high-impact refereed journals.

Dr. Zou was a recipient of the Alexander von Humboldt Research Fellowship, National Outstanding Expert in Science & Technology of China, the Nomination Award for the National Excellent Doctoral Dissertation of China, the Science & Technology Award for Young Scientist of Sichuan Province, China, and the Outstanding Reviewer of *Optics Communications*. He currently serves as an Associate Editor of *IEEE Journal of Quantum Electronics*, and once as the Editor and the Coordinator for a Focus Issue on Microwave Photonics in *IEEE Journal of Quantum Electronics*, and the leading Guest Editor for a Special Section on Microwave Photonics in *Optical Engineering*.

**Wei Pan** is a currently Full Professor and the Dean of the School of Information Science and Technology, Southwest Jiaotong University, China. His research interests include semiconductor lasers, nonlinear dynamic systems, and optical communications. He has published over 200 papers in high-impact refereed journals.

Dr. Pan was appointed as the steering and consultancy expert of the highest level national project (973 Project) of China. He was also awarded as the Academic and Technology Leader of Sichuan province, China.

**Lianshan Yan** is currently a Full Professor and the Director of Center for Information Photonics and Communications in Southwest Jiaotong University, China.

Dr. Yan received the IEEE Photonics Society Distinguished Lecturer Award for 2011-2013 and the IEEE LEOS Graduate Fellowship in 2002. He currently serves as the Chair of the Fiber Optics Technology Technical Group, the Optical Society of America (OSA). He served as the co-chair or the TPC member of more than 20 international conferences, including the Optical Fiber Communication Conference and Exposition (OFC, 2013-), the European Conference on Optical Communication (ECOC, 2013-), the Asia Communications and Photonics Conference (ACP, 2010-2012), etc. He once served as an Associate Editor of *IEEE Photonics Journal*.

**Shilong Pan** is currently a Full Professor and the Executive Director of the Key Laboratory of Radar Imaging and Microwave Photonics (Ministry of Education), Nanjing University of Aeronautics and Astronautics. From 2008 to 2010, he was a "Vision 2010" Postdoctoral Research Fellow in the Microwave Photonics Research Laboratory, University of Ottawa, Canada. His research has focused on microwave photonics, which includes optical generation and processing of microwave signals, photonic microwave measurement, and integrated microwave photonics. He has authored or co-authored more than 130 papers in peer-reviewed journals.

Dr. Pan was selected to receive an OSA outstanding Reviewer Award in 2015. He is currently a Topical Editor for *Chinese Optics Letters*. He was the Chair of numerous international conferences and workshops, including the TPC Chair of the International Conference on Optical Communications and Networks in 2015, the TPC Chair of the high-speed and broadband wireless technologies subcommittee of the IEEE Radio Wireless Symposium in 2013, 2014, and 2016, and the Chair of the microwave photonics for broadband measurement workshop of International Microwave Symposium in 2015.